\documentstyle[floats,graphicx,twocolumn,aps]{revtex}
\begin{document}
\draft

%\preprint{HEP/123-qed}
\title{Measurements of the Magnetic Field Dependence of ${\bf \lambda}$ in 
${\bf YBa_2Cu_3O_{6.95}}$:  Results as a Function of Temperature and 
Field Orientation}
\author{C.P. Bidinosti, W.N. Hardy, D.A. Bonn and Ruixing Liang}
\address{Department of Physics and Astronomy, University of 
British Columbia, Vancouver,
         B.C., Canada  V6T 1Z1}
\date{\today}
\maketitle

\begin{abstract}
We present measurements of the magnetic field dependence of the
penetration depth $\rm \lambda(H)$ for untwinned
${\rm YBa_2Cu_3O_{6.95}}$ for temperatures from 1.2 to 70 K in dc fields 
up to 42 gauss and 
directions $0^{\circ}, \pm 45^{\circ}$ and 
$90^{\circ}$ with respect to the crystal b-axis.  The experiment uses 
an ac susceptometer with fields applied parallel to the ab-plane of thin 
platelet samples.  The resolution is about 0.15 ${\rm \AA}$ in zero dc
field, degrading to 0.2 or 0.3 ${\rm  \AA}$ at the higher fields. 
At low temperatures the field dependencies are essentially linear in H,
ranging from 0.04 ${\rm \AA}/$gauss for $\Delta \lambda_{a}$ to 0.10
${\rm \AA}/$gauss for $\Delta \lambda_{b}$, values comparable to the 
  ${\rm T=0}$ Yip and Sauls prediction for a d-wave superconductor.  
However, the systematics versus temperature and orientation 
do not agree with the d-wave scenario probably due, in part,
to residual sample problems.
\end{abstract}
\pacs{74.25.Nf, 74.25.Ha, 74.72.Bk, 74.90.+n}

\narrowtext

When found, a successful theory for the unconventional superconductivity in
the HiTc cuprates will have to encorporate a detailed knowledge
of the structure of the pairing
state.  It is for this reason, that experiments studying the energy gap
and its
symmetry, and therefore the symmetry of the order parameter and pairing
state, have always been of great importance.  However, for sometime after
the discovery of the HTSC's, experiments probing the gap produced 
conflicting
results and  left much debate as to whether or not the pairing state had
the conventional s-wave symmetry of BCS theory.
  Yip and Sauls \cite{ys}  proposed experiments based on
the nonlinear effects of an applied magnetic field on the
Meissner state
supercurrent that would differentiate between s-wave and d-wave
superconductors (the d-wave state with 
${d_{x^{2}-y^{2}}}$ symmetry having long been the favoured
alternative to s-wave for the cuprates.)
Today, however, this debate is largely settled despite the theory of the
nonlinear Meissner effect remaining grossly under tested.  
Mounting evidence from several other experiments 
\cite{kirtley,vanharligen,arpes} has left little
doubt that for HTSC's the pairing state is essentially d-wave, exhibiting
a predominantly ${d_{x^{2}-y^{2}}}$ symmetry.

Nevertheless, the work of Yip and Sauls and that of others
\cite{xys,sv,iv1,iv2,betouras} is still very much worth investigation.
The nonlinear Meissner effect, as predicted, will lead to both a field
dependent penetration depth as well
as a magnetization that can contain a component transverse to the applied
field.  Furthermore, both quantities should  show an anisotropy
dependent upon field direction with respect to the nodes of a d-wave gap.
In particular, for $\lambda({\rm H})$ theory
predicts that at sufficiently low temperatures there is a linear field 
dependence of the form 
\begin{equation}
\lambda({\rm T,H})- \lambda({\rm T,0})= \Delta \lambda({\rm
H}) = \lambda({\rm T})
\alpha \left|{\rm H}\right|/{\rm H_{o}(T)},
\label{eq:1}
\end{equation}
where $\alpha$ is equal to unity for fields along a node and $1/\sqrt{2}$ 
when fields are along an antinode; 
the quantity ${\rm H_{o}(T)}$ is a characteristic field of order the
 thermodynamic critical field \cite{xys}.  In contrast, a superconductor
with a conventional gap would show an  ${\rm H^{2}}$ dependence
of $\Delta \lambda$ with a thermally activated prefactor and 
with no anisotropy with respect to field direction.  
It is the prospect of using
these measurements as a probe
to determine the positions of the nodes that is now probably the most exciting 
aspect of this theory. 

Unfortunately, experiment has lagged far behind the theory and 
{\it node spectroscopy} of this kind has
never been successfully achieved.  Experimenters face the difficult task of 
searching for a  very small 
magnetic field effect hidden in a background of extrinsic field dependencies
and other inherent sample 
anisotropies (for example, geometry, twinning, orthorombicity), all the while
being restricted to the Meissner 
state, which limits the field range one can study.
Early measurements of 
$\Delta \lambda({\rm H})$ in ${\rm YBa_2Cu_3O_{7-\delta}}$
 \cite{sridhar,maeda} were limited to  
a resolution of $ \sim$ 20 $\rm \AA$ (two orders of magnitude less than the 
method to be described in this paper), and neither investigated the
anisotropy as a function of field in the ab-plane. 
The Minnesota group has concentrated on the angular dependence of the 
transverse magnetization $\rm M_{\perp}$.  Results for 
${\rm LuBa_{2}Cu_{3}O_{7-\delta}}$ 
did not support a pure d-wave pairing state, but measurements were
dominated by the geometric demagnetization factor and trapped 
flux \cite{buan}. Subsequently, efforts have been made
to alleviate the problems associated with sample
geometry\cite{bhattacharya}.
 One cannot easily compare the sensitivity of the  
$\rm M_{\perp}$ measurements to our $\Delta \lambda({\rm H})$ measurements, 
but ultimately results from both types of experiment should complement
each other. It is the aim of this paper to show that we  have made
substantial advances in testing the
nonlinear Meissner effect, firstly by the development of a high resolution 
technique for measuring $\Delta \lambda({\rm H})$ for fields in the ab-plane,
and secondly by the demonstration that in ${\rm YBa_2Cu_3O_{6.95}}$ the 
effect of magnetic fields on $\lambda$ is much smaller than that suggested
by previous $\Delta \lambda({\rm H})$ experiments. 

Superconducting microwave cavity perturbation has proven to be a very
successful technique 
for measuring small temperature dependencies in $\lambda$ of
superconductors in the Meissner state \cite{hardy,lee,broun}. However,
this method is impractical for field dependent measurements.
The application of a dc field inside the superconducting cavity is not
possible, and the power dependence of the cavity severely inhibits 
studies as a function of microwave intensity.  Here, an ac susceptometer
has been designed that has very near the resolution of the microwave
technique for variations in temperature ($\delta \lambda({\rm T}) \sim$
0.15 ${\rm \AA}$), but can also be used to make field dependent measurements 
with only slightly less resolution ($\delta \lambda({\rm H}) \sim$ 0.2 to
0.3 ${\rm \AA}$).  

\begin{figure}
\centering
%\vskip 1pc
\includegraphics[keepaspectratio,width=3.2in]{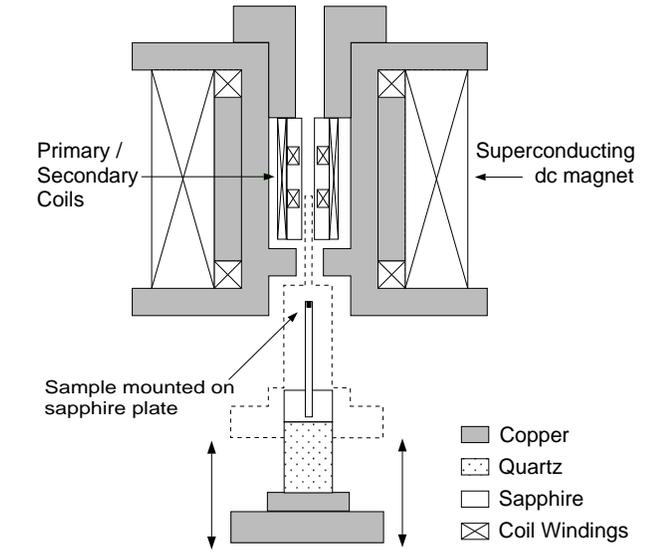}
\caption{The ac susceptometer:  constructed from only
materials with very small magnetic susceptibilities. The sample is shown
extracted from the
pick-up coil; the dashed outline shows the sample in the measurement
position.}
\label{fig:acsus}
\end{figure}

Such precise measurements of $\Delta \lambda({\rm T,H})$ require
the sample to be very well isolated from the
background.  Thermal isolation is achieved by mounting the sample
on a sapphire cold finger in vacuum \cite{sridhar2,rubin}, a technique that is
widely in use.  Magnetic isolation is an extremely difficult 
condition to satisfy, since it is not possible to avoid  exposing certain 
parts of the apparatus to the field as well; 
Figure~\ref{fig:acsus} exhibits this problem quite clearly.  
The obvious approach is to use only `nonmagnetic' materials in the
construction of the apparatus, but at the level of sensitivity
required here
this did not sufficiently suppress a field dependent 
background signal.  To overcome this problem, a custom made retractable
sample holder was developed (see also Figure~\ref{fig:acsus}).  Measurements 
of $\Delta \lambda({\rm H})$
 are made with a swept dc field; the amplitude of the ac field is fixed.
The background signal 
 is determined by 
extracting the sample from the susceptometer and repeating the same dc
field sweep. 
Tests with only the sapphire sample plate and the equivalent amount of vacuum 
grease to hold a sample to the plate show no regular field dependence once the
 background is subtracted out. The peak values of the residual signal from
this test correspond to a change of approximately 0.2
to 0.3 ${\rm \AA}$ in penetration depth for a typical sized 
${\rm YBa_2Cu_3O_{7-\delta}}$ sample.  We consider this to be our 
resolution 
in $\Delta \lambda({\rm H})$.

The susceptometer operates with a 12 kHz ac field supplied by the primary coil.
The output voltage from the secondary coils is proportional to the sample
magnetization.  To relate the change in voltage $\Delta v$ to a change in
penetration depth $\Delta\lambda$ for a thin plate superconductor, we
determine a calibration constant ${\rm k}=\Delta\lambda /\Delta v$ 
valid in the limit $2\lambda << \rm t$, the sample thickness.
(When  $2\lambda \sim \rm t$ such a simple relation between  $\Delta v$
and $\Delta\lambda$ does not exist.)
This is done by heating the sample from the base temperature
${\rm T_{b}} = 1.2$ or 4.2 K to above ${\rm T_c}$; at 12 kHz $\rm t <<
\delta$, the normal state skin depth, so the resulting change in
voltage going from
the superconducting to the normal state $v_{\rm sn}$ can be related to
the change in the effective superconducting volume of the sample
 $\simeq {\rm V}-2{\rm A}\lambda({\rm T_{b})}$,
where A is the sample area and V its volume both at $\rm T_{b}$.  
The calibration constant k is then  
\begin{equation}
{\rm k} = ({\rm t}-2\lambda({\rm T_{b}}))/2v_{{\rm sn}}
\simeq {\rm t}/2v_{{\rm sn}}
\end{equation}

\begin{figure}
\centering
%\vskip 1pc
\includegraphics[keepaspectratio,width=3.4in]{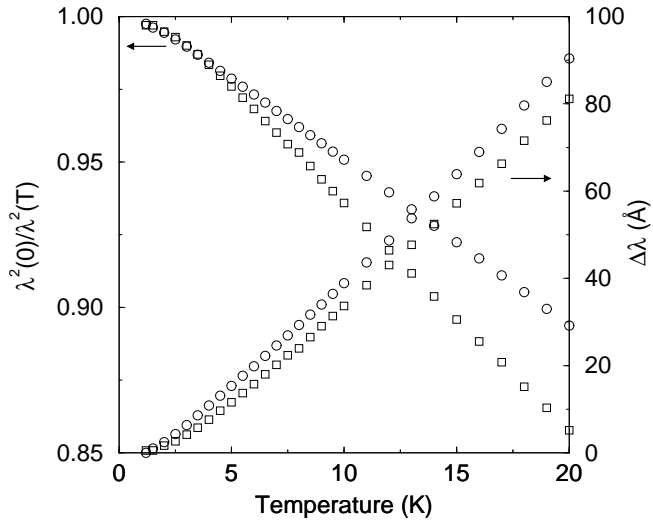}
\caption{$\Delta\lambda(\rm T)$ and the superfluid
fraction $\lambda^{2}(0)/\lambda^{2}(\rm T)$ in zero dc field for
a-(circles) and
 b-(squares) directions.}
\label{fig:dlt}
\end{figure}

Measurements were  made on a detwinned single crystal of 
${\rm YBa_2Cu_3O_{6.95}}$ grown in an
yttria-stabilized crucible \cite{ruixing}.  The sample had dimensions
$0.98:1.89:0.056$ in mm.  It had a  ${\rm T_c}= 91.8$ K, also measured
using the 
ac susceptometer, and exhibited the characteristic linear
$\lambda({\rm T})$ behavior of a high quality cuprate superconductor at 
low temperatures.  Figure~\ref{fig:dlt} shows a graph of 
$\Delta\lambda_{a}$ and $\Delta\lambda_{b}$ as a function of temperature.  
For both quantities the linear $\Delta \lambda({\rm T})$  
crosses over to higher powers in T below 4.5 K. This 
crossover is due to an impurity level that is typical of crystals grown in
yttria-stabilized crucibles \cite{saeid_bzo} and is not believed to be
large enough to mask the nonlinear Meissner effect \cite{xys}.
  
The crystal was initially mounted so the fields were parallel to the b-axis 
in the ab-plane.  It was subsequently  rotated in three $45^{\circ}$
increments around the c-axis.  An extensive set of data was taken for each
crystal orientation.  Measurements  of $\Delta\lambda({\rm T})$ were made
in zero dc field from 1.2 K to 20 K; the ac drive field had a peak
amplitude of $\sim$ 7 gauss. Measurements of $\Delta\lambda({\rm H})$  
were made for a series of fixed temperatures from 1.2~K to 70~K with the
same amplitude for the drive field.  The dc field was stepped in even
increments through a loop of $\pm$ 42 gauss.  Data was collected for 10
successive loops and averaged.  Similar runs were done with the sample
extracted from the susceptometer in order to determine the background. A 
background run was not taken after every data run and often the same
background was used for several data sets.  
However, experimental conditions were never allowed to change markedly
before a background run was performed, and
careful attention to the signal gain of the system showed that this time 
saving measure did not introduce any noticeable error into our analysis.

As mentioned, the results for  $\Delta\lambda_{a}(\rm T)$ and  
$\Delta\lambda_{b}(\rm T)$ are 
shown in Figure~\ref{fig:dlt}.  Results for the ac field $45^{\circ}$ to the 
crystal axes (data not shown)
also exhibit the linear T dependence along with the crossover to higher
powers below about 4.5 K.  Linear fits to $\Delta \lambda({\rm T})$ 
above 4.5 K give slopes of 
5.0 ${\rm \AA}/$K and 4.6 ${\rm \AA}/$K for the a- and b-directions 
respectively, and 4.7 to 4.8 ${\rm \AA}/$K for the $45^{\circ}$
orientations.  The magnitudes of these slopes are typical for ${\rm
YBa_2Cu_3O_{6.95}}$, as is the result that $\Delta\lambda_{a}({\rm
T})/\Delta{\rm T} > \Delta\lambda_{b}(\rm T)/\Delta{\rm T}$.
Furthermore, as expected, the values for the $45^{\circ}$ orientations 
are averages of the values for the a- and b-directions.  These results give 
us confidence that the zero field electrodynamics of our crystal are no 
different from previously published data and furthermore that we have 
not misidentified the a- and b-axes.

\begin{figure}
\centering
%\vskip 1pc
\includegraphics[keepaspectratio,width=3.3in]{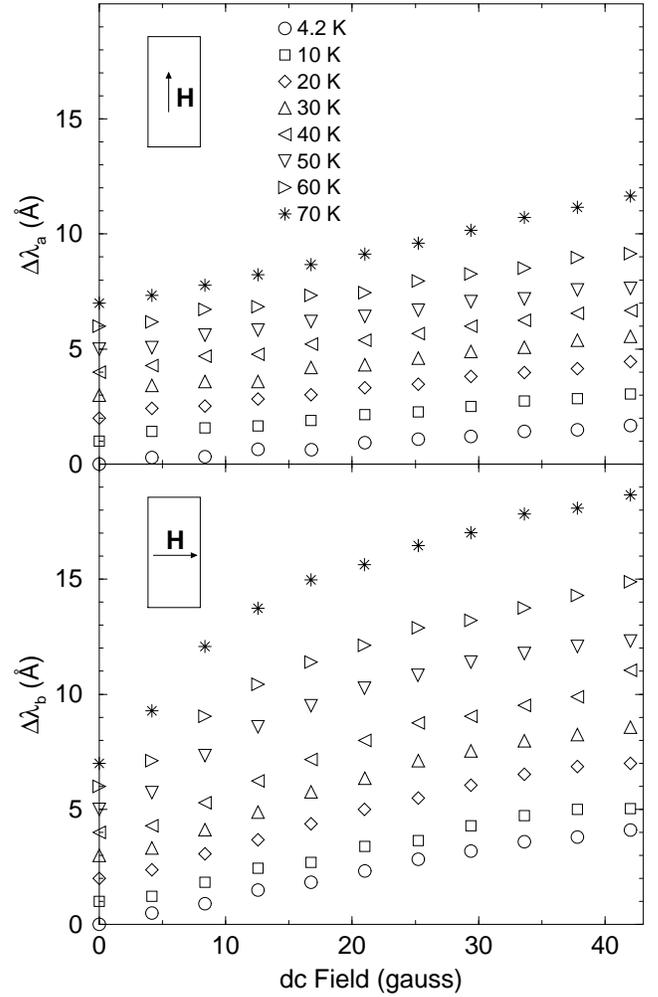}
\caption{$\Delta\lambda$ as a function of H for fields
parallel (top) and perpendicular (bottom) to  the crystal b-axis. }
\label{fig:dlh1}
\end{figure}

\begin{figure}
\centering
%\vskip 1pc
\includegraphics[keepaspectratio,width=3.3in]{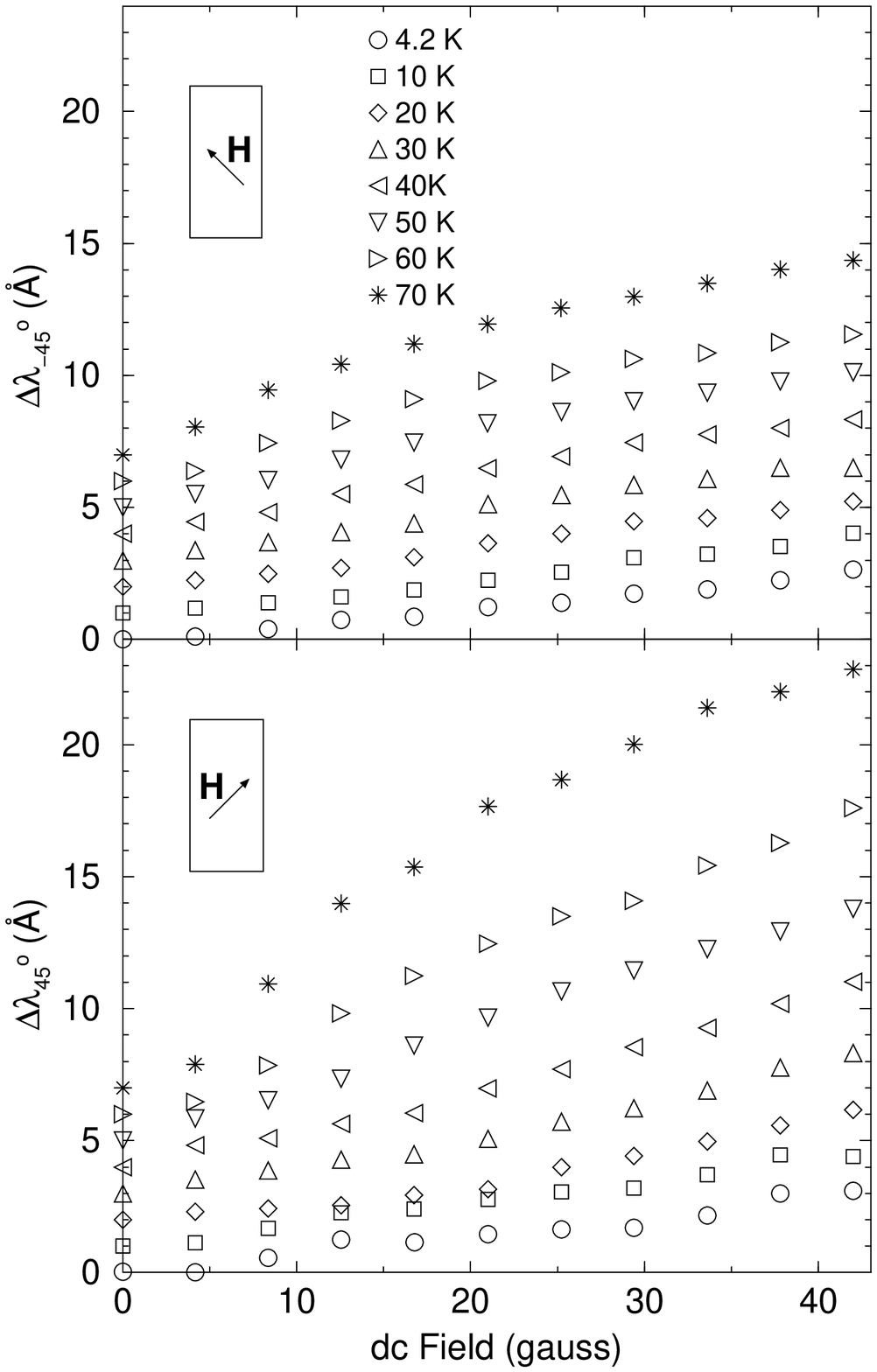}
\caption{$\Delta\lambda$ as a function of H for fields $\pm 45^{\circ}$ to
the
crystal axes. }
\label{fig:dlh2}
\end{figure}

The results for $\Delta \lambda({\rm H})$ are shown for the a- and 
b-directions in 
Figure~\ref{fig:dlh1}. The results for  $\pm 45^{\circ}$, shown in
Figure~\ref{fig:dlh2}, should be identical for a rectangular shaped 
orthorhombic crystal; clearly they are not.  In each
graph the data sets have been separated by 1 ${\rm \AA}$ at zero field for the 
purpose of clarity.  The 1.2 K data has not been plotted, as it is not 
distinguishable (within resolution) from the 4.2 K results.  All other data 
from 4.2 K to 70 K is presented, and it is clear from this that the effect
of field on 
$\Delta \lambda$ increases with temperature.  This temperature dependence
is clearly not exponential, which is consistent with the belief that the
energy gap in ${\rm YBa_2Cu_3O_{6.95}}$
 is not s-wave. The strongest test of a d-wave gap
within the theory of the nonlinear Meissner effect is the linear field
dependence in $\Delta\lambda$.  This is seen for $\Delta\lambda_{a}$ over
most of the
temperature range studied.  For the other sample orientations, it is only
at low temperatures where the field dependence is 
essentially linear.  Linear fits to the 4.2 and 10 K data give slopes of 
about 0.04 and 0.10~${\rm \AA}/$gauss for the a- and b-directions
respectively, and about 0.07~${\rm \AA}/$gauss for the  $\pm 45^{\circ}$
directions.  In contrast, measurements at 12 K by Maeda et al.\cite{maeda}  
(measured for fields perpendicular to the ab-plane) gave a linear field 
dependence of
$\sim$ 8~${\rm \AA}/$gauss, roughly 100 times larger 
then our result.  In even greater contrast is the quadratic field dependence 
seen by Sridhar et al. \cite{sridhar} in 1989, but this is most likely
attributable to the general unavailability of high quality crystals at
that time.  

Although this experiment has improved the sensitivity of 
$\Delta \lambda({\rm H})$ 
measurements by nearly two orders of magnitude and shown that the nonlinear
effects are much smaller than previously reported, it is clear that we have
not yet reached the intrinsic values, at least in the $\pm 45^{\circ}$
 and b-directions.  First of all, the fact that the $+45^{\circ}$ and
$-45^{\circ}$
measurements do not agree shows us that the results do not follow the
symmetry of a rectangular shaped, orthorhombic crystal.  Possible mechanisms
for this result are remnant microtwins from the detwinning process, or 
sharp-edge effects that are not symmetrically distributed.  The much
larger
slope for the b-direction is also surprising, and possibly due to some 
nonlinear effect associated with the Cu-O chains. However the a-axis results,
being the most linear in H and having the weakest temperature effects, are 
likely closest to the intrinsic values. 
 
Xu, Yip and Sauls (XYS)~\cite{xys} define the characteristic field 
$\rm H_{o}=(3v_{o}c)/(2e\lambda)$
where $\rm v_{o} \sim \Delta_{o}/v_{f}$, the gap maximum divided by the Fermi
velocity; they estimate $\rm H_{o}$ to be approximately 2.5 T.  To extract
$\rm H_{o}$ from our a-direction data we rearrange Equation~\ref{eq:1} as
$\rm H_{o} =
\lambda_{a}\alpha_{exp}({\rm H}/\Delta\lambda_{a})$, where the quantity in 
brackets is the reciprocal of the slope to our linear fit given above and 
$\rm \alpha_{exp} = 1/\sqrt{2}$ \cite{alpha}. Using 
$\lambda_{a}=1600$~${\rm \AA}$ \cite{lambdas} we find 
a characteristic field of 2.8~T for temperatures in the 
range 1.2 to 10~K.
 However, as discussed in XYS~\cite{xys}, thermal effects 
cause the ${\rm T=0}$ linear field dependence to crossover to quadratic; the 
crossover field being given by $\rm H_{T}=(T/\Delta_{o})H_{o}$ equals 60 gauss 
at our lowest temperature of 1.2~K.  Thus the full Yip and Sauls theory 
predicts a weaker, quadratic dependence of $\rm \Delta\lambda(H)$ in our 
temperature and field range, and the above agreement is probably accidental.

In summary, we have presented measurements of 
$\Delta \lambda({\rm H})$ for an untwinned crystal 
of ${\rm YBa_2Cu_3O_{6.95}}$ made as a function of temperature 
and, for the first time, as a
function of field orientation in the ab-plane.  The resolution of this
experiment is $\sim$ 100 times greater than any previous experiment
measuring $\Delta \lambda({\rm H})$.  
At low temperatures a linear field dependence in$\Delta\lambda(\rm H)$ is 
observed, which while apparently
agreeing with the ${\rm T=0}$ limit of the Yip and Sauls 
theory is not in agreement when thermal effects are
taken into account. Also, there is an anisotropy
in $\Delta \lambda({\rm H})$ for fields $\pm 45^{\circ}$ to the 
b-axis that can only be explained by some mechanism that breaks the crystal 
symmetry;  we suggest remnant twinning, or unsymmetrical edge effects.

\acknowledgments
The authors acknowledge the assistance of S. Kamal and P. Dosanjh.
This work was supported by the National Science and
 Engineering Research Council of Canada and the Canadian Institute
 for Advanced Research. DAB gratefully acknowledges support from
 the Sloan Foundation.



\begin{references}
\bibitem{alpha} The coefficient $\rm \alpha_{exp}$ may differ from the 
value of $\alpha$ given by XYS depending on the definition of 
$\lambda$ and the measurement technique.
In reference~\cite{xys}, $\lambda$ is  defined by the initial 
rate of decay of the magnetic field inside the sample.  
For a volume exclusion experiment, such as ours, 
a more appropriate definition
is $\lambda=\int_{0}^{\infty}\rm H(z)/H(0)\,dz$, which
reduces $\Delta\lambda(\rm H)$ by a factor of 2. On the other hand, 
 an ac measurement will result in a $\Delta\lambda(\rm H)$ twice 
that of a dc measurement.  Thus, for our ac susceptibility measurements, the 
two factors cancel and the appropriate $\rm \alpha_{exp}$ is numerically
equal to the XYS value.
\end{references}
\end{document}